# Identifying vital edges in Chinese air route network via memetic algorithm

Wen-Bo Du[a,b], Bo-Yuan Liang[a,b], Gang Yan[c], Oriol Lordan[d], Xian-Bin Cao[a,b]*

[a]*School of Electronic and Information Engineering, BeihangUniversity, Beijing 100191, P.R.China*
[b]*Beijing Key Laboratory for Network-based Cooperative Air Traffic Management, Beijing 100191, P.R.China*
[c] *School of Physics Science and Engineering, Tongji University, Shanghai 200092, P. R. China*
[d]*Universitat Politècnica de Catalunya-BarcelonaTech, C/Colom no. 11, Terrassa 08222, Spain*

**Abstract**

  Due to its rapid development in the past decade, air transportation system has attracted considerable research attention from diverse communities. While most of the previous studies focused on airline networks, here we systematically explore the robustness of the Chinese air route network, and identify the vital edges which form the backbone of Chinese air transportation system. Specifically, we employ a memetic algorithm to minimize the network robustness after removing certain edges hence the solution of this model is the set of vital edges. Counterintuitively, our results show that the most vital edges are not necessarily the edges of highest topological importance, for which we provide an extensive explanation from the microscope of view. Our findings also offer new insights to understanding and optimizing other real-world network systems.

*Keywords:*  vital edges; air route network; memetic optimization.

## 1. Introduction

   With the increasing people and goods transport demand during the accelerating globalization process, the air transportation system plays a more and more important role than ever before due to its high-speed and high-security advantages. For example, the air transport volume of China grows at an average annual speed of over 10% in the past decades, and now it possesses over one seventh of the total comprehensive transport volume (including roadways, railways, shipping and air transport), which was only 7.9% in 2000. Hence the air transportation system has been drawing much attention from different research communities. One of the most interesting directions is to analyze the structure and function of air transportation systems within the framework of complex network theory.
   The air transportation system can be represented as a network, in which nodes denote airport and an edge will be created if there is a direct flight between two airports. In the vast majority of previous literature, the air transport network (ATN) was primarily classified into two scales: worldwide and national.
   For the worldwide scale, Amaral et al. firstly found worldwide ATN is a small-world network with a power-law degree distribution, and the highest-degree airport is not necessarily the most central node, prompting them to propose a network model where both geographical and political factors are taken into account[1-2]. Barrat et al. investigated the worldwide ATN from a perspective of complex weighted networks and found the nonlinear positive correlation between flight flow and topology properties[3-4]. They proposed a weighted network model, enlightening the understanding of weighted feature of complex systems. Verma et al. decomposed the worldwide ATN into three distinct layers via k-core decomposition and found that this network is robust to the removal of long distance edges, but fragile to the disconnectivity of short and apparently insignificant edges[5-6].
   For the national scale, ATNs of several major nations, such as US, Brazil, India and China, are extensively studied[3, 7-11], and the national ATNs usually exhibit different features from the worldwide ATN. Gautreau et al. studied US ATN during 1990–2000[3]. A remarkable result they presented is that although most statistical properties are stationary, an intense activity takes place at the local level. Fleurquin et al. proposed a delay propagation model via quantifying the network congestion for US ATN, revealing that even under normal operating condition the systemic instability risk is non-negligible[11]. Rocha investigated the Brazilian ATN during 1995-2006, and found that it shrank in topology but grew in traffic volume [7]. Bagler et al. studied the Indian ATN, and found its signature of hierarchy feature[12]. As the most active economy, the Chinese aviation industry ranks second to US in the past decade and keeps a high increase rate. Consequently, Chinese ATN attracts continuous attention in different aspects, from topology to dynamics and evolution [8-10, 13-14], one of which is to investigate the backbone of ATN, the air route network (ARN).

---

∗Corresponding author. Tel.: +86-10-82314318
*E-mail address:* xbcao@buaa.edu.cn

ATN is actually a logic network with Origin-Destination (OD) relationships. In real air traffic operation, a flight does not straightly fly from departure airport to landing airport, but along some air route waypoints. ARN consists of air route waypoints and connections between them. In 2012, Cai et al. firstly investigated the Chinese ARN[15] and found that the degree distribution of Chinese ARN is homogeneous but the traffic flow on it is rather heterogeneous. Vitali et al then investigated the horizontal deviation and delays in Italian ARN[16]. The analysis of ARN is quite a novelty in the literature. However, the network robustness, which is an important issue for infrastructure systems[17] and has been extensively studied in ATN [18-21], is still rare in ARN. In the typical network robustness model, edges are removed by different targeted attack strategies and the size of giant component estimates the robustness of the network[30]. When a small amount of edges are removed, the size of giant component is of a very small change. In this paper, we focus on identifying the vital edges in Chinese ARN by examining the robustness of the new network after removing an edge set via memetic optimization. Remarkably, we find that the most vital edges are not necessarily the edges of highest topological importance.

The rest of paper is organized as follows. In the next section we demonstrate Chinese air route network and its basic properties. Section 3 describes the optimization model and the memetic algorithm. Section 4 presents the simulation results and corresponding analysis. Finally, the paper is concluded in Section 5.

## 2. Chinese air route network

The latest data of the Chinese air route network are provided by the Air Traffic Management Bureau (ATMB) of China. In the Chinese ARN, airports or air route waypoints are nodes and edges are represented by the air route segments. An air route waypoint is a navigation marker which keeps the pilots informed about the desired track. In the air transportation system, the flights will fly along the air route waypoints, but not directly fly from one airport to another (Fig.1). Figure 1 is an illustration of ARN, where airlines are depicted by the dotted line and air route segments are denoted by the solid line. Figure 2 shows the structure of the Chinese ARN, which contains $N = 1,499$ nodes and $M = 2,242$ edges.

In ref. [15], the authors found that the topology structure of the Chinese ARN is homogeneous, yet its distribution of flight flow is quite heterogeneous. If we compare the Chinese ATN with the Chinese ARN we found significant differences. On one hand, the Chinese ATN is a typical small-world with low average shortest path length and large clustering coefficient. On the other hand, the Chinese ARN is not a small-world network due its low clustering coefficient, large average shortest path length and exponential spatial distance distribution.

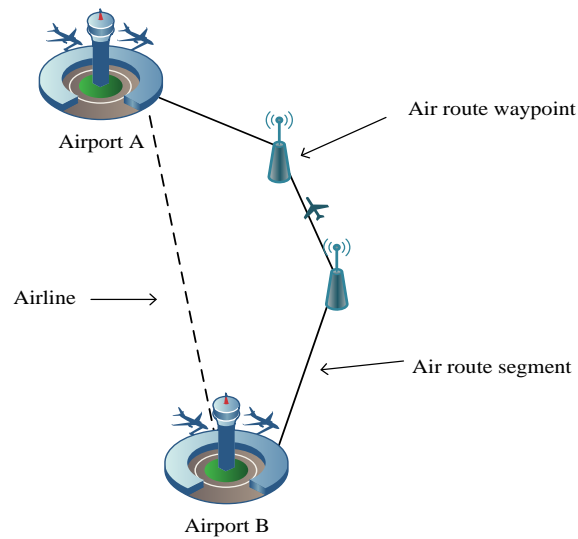

Fig. 1    An illustration of ARN.

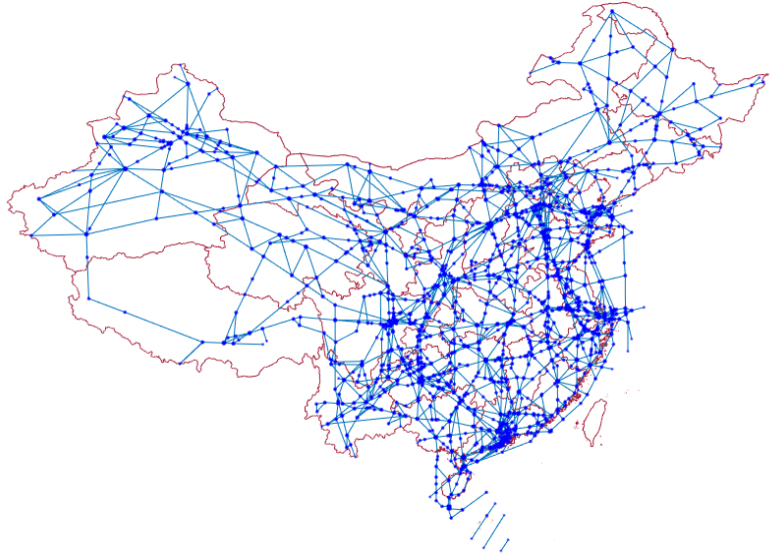

Fig.2 The structure of the Chinese ARN. Here the Chinese ARN contains $N = 1,499$ nodes and $M = 2,242$ edges.

## 3. Model

### 3.1. Optimization Model

The static robustness of complex networks has been extensively studied in the past decades. In ref [22], it is quantified by the relative size of the largest connected component $G = N'/N$, where $N$ is the total number of nodes in initial network and $N'$ is the number of nodes in largest component after attack. The larger value of $G$ represents a more robust network. Based on the largest connected component, Schneider et al. proposed a measure $R$ to evaluate the robustness against targeted attack on nodes[17].

$$R = \frac{1}{N}\sum_{Q=1}^{N} s(Q)$$

where $s(Q)$ is the fraction of nodes in largest component after removing $Q$ nodes. For calculating the robustness of a network we will follow a degree adaptive strategy: the highest degree nodes will be systematically remove one by one. It is a more comprehensive measure of network robustness. Obviously, a network with higher $R$ has a stronger resistance to targeted attacks.

In the Chinese ARN, the closure of air route segment will decrease the connectivity of the whole network. It is of great significance to identify vital edges that lead to the vulnerability of ARN. In ref [23], Freeman proposed a global metric edge-betweenness to measure the importance of an edge, which can identify influential edges effectively. It is defined as follows,

$$B_E(i) = \sum_{j,k \in V, j \neq k} \frac{n_{jk}(i)}{n_{jk}}$$

where $V$ is the set of nodes, and $n_{jk}(i)$ denotes the number of shortest paths from $j$ to $k$. $n_{jk}(i)$ denotes the number of shortest paths from $j$ to $k$ that via edge $i$.

In this work, we formulate a combinational optimization problem to identify the vital edges within network robust model in the Chinese ARN. The objective is to minimize the network robustness after removing certain edges, i.e. closing certain air route segments. Therefore, these edges play an important role in maintaining network robustness. The optimization model is formulated as follows,

$$\min \ R'(x)$$
$$x = [e(1), e(2), \ldots, e(V)]$$
$$s.t. \sum_{i=1}^{V} x_i = cost$$

where $V$ is the total number of edges in the network and $x$ is a $V$ dimensional binary variable, $e(k) \in \{0, 1\}$ ($e(k)$ represents the $k_{th}$ edge in the network). $e(k) = 1$ represents the edge $e(k)$ is removed, otherwise it remains in the network. The total number of removed edges is certain and denoted as cost (*C*). So, we can identify critical edges in the network minimizing its robustness.

*3.2. Memetic Algorithm*

For solving this optimization model we will use the memetic algorithm (MA), a useful tool for dealing with large-scale combinational problem[24-27]. Coming from the concept of meme, MA is defined as a part of local improvement in the process of cultural evolution. It is a hybrid metaheuristic of global search and heuristic local search with three operations: crossover, local search and tournament selection.

**Initialization**

In MA, the population is composed of $X_n$ individuals. Each individual $x$ represents a scheme of removing edges in the network and was generated randomly.

**Crossover**

The crossover operator works on two parent individuals and can search in a large area. Suppose that $x_{p1}$ and $x_{p2}$ are two parent individuals, and $x_{c1}$ and $x_{c2}$ are two child individuals. First, we assign $x_{p1}$ to $x_{c1}$ and $x_{p2}$ to $x_{c2}$ and obtain the following sets of edges.

$$E_c = \{e | e \in x_{c1}, e \in x_{c2}\}$$
$$E_{\overline{c1}} = x_{c1} - E_c$$
$$E_{\overline{c2}} = x_{c2} - E_c$$

In fact, $E_c$ is the set of common edges of $x_{c1}$ and $x_{c2}$. $E_{\overline{c1}}$ and $E_{\overline{c2}}$ are the set of edges of the network after removing the common edges of $x_{c1}$ and $x_{c2}$. That is, $E_{\overline{c1}}$ and $E_{\overline{c2}}$ have the same number of edges but completely different. Then, for each pair of edges in $E_{\overline{c1}}$ and $E_{\overline{c2}}$ we conduct the following operations with the probability $P_c$.

$$x_{c1} = x_{c1} - E_{\overline{c1}}[i] + E_{\overline{c2}}[i]$$
$$x_{c2} = x_{c2} - E_{\overline{c2}}[i] + E_{\overline{c1}}[i]$$

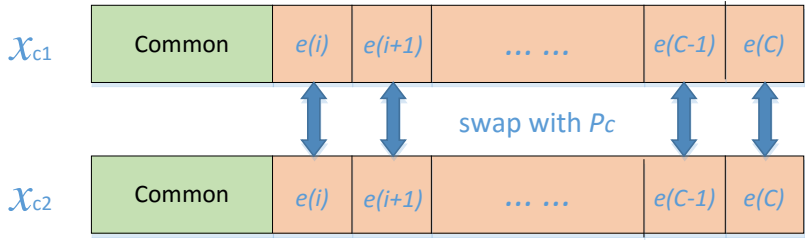

Fig.3　An illustration of crossover process.

In summary, only the set of non-common edges we want to remove will be swapped between $x_{c1}$ and $x_{c2}$ (Fig. 3).

**Local Search**

The local search operator is an important part in MA that can accelerate the convergence speed. Based on previous edge importance evaluations[28-29], we adopted a local search in the direction of removing more important edges. We first select an individual from parent and child population using the roulette wheel selection based on their fitness. Then for each edge of the selected individual we conduct a local search with probability $P_l$. For example, edge $e_{ij}$ ($e_{ij}$ represents the edge between node *i* and node *j*) will mutate to a randomly selected existing edge $e_{lm}$ but not in its individual. This mutation will be accepted when the following formula is satisfied.

$$\mu \cdot |k_l - k_m| - |k_i - k_j| > 0$$

where $\mu$ is a formula parameter in the range [0,1] and $k_l$, $k_m$, $k_i$ and $k_j$ represents the degree of nodes *l*, *m*, *i* and *j* respectively.

**Tournament selection**

In this part, two individuals are respectively chosen from parent population and child population to run a tournament for the tournament selection. The population with the best fitness is selected for the next generation, and the total number of generation is $P_m$.

To conclude, the pseudocode of MA proposed is presented as follows.

Table 1 The pseudocode of MA

```
Begin
Initialize P_n individuals with randomly selected existing edges in network.
While g < P_m
   Repeat
       Randomly selected two individual x_pi and x_pj that have not been selected.
       If rd < p_c
           Assign parent x_pi and x_pj to child x_ci and x_cj.
           (x_ci, x_cj) ←Crossover (x_pi, x_pj).
       end if
   Until (all individuals have been selected)
   For i = 1 to 2P_n
       select an individual x_l using the roulette wheel selection based on the fitness.
       For each edge e_ij in individual x_l
           if rd < p_l
               mutate with a new randomly selected edge e_lm but not in x_l.
               if μ·|k_l − k_m| − |k_i − k_j| > 0
                   Accept the local search.
               end if
           end if
       end for
   Calculate the robustness of each individual in parent and child.
   P_next ← 2-Tournament Selection (P_parent, P_child);
end while;
End
```

## 4. Result

Here, in order to identify the vital edges, we examine the robustness of the new networks after removing edges via memetic algorithm (MA), and compare it with the highest edge-betweenness adaptive strategy ($B_EAS$). To demonstrate the universality of our method, the experiments are carried out on both Chinese ARN and the general Barabási- Albert (BA) scale-free network. The Chinese ARN has 1,499 nodes and 2,242 edges. The BA scale-free network is generated with $m_0$ nodes and a new node is added with $m$ edges at each time step, which connect the new node with $m$ different existing nodes. Here, it is set that $m_0 = 2$ and $m = 2$ and the BA network is of 1,000 nodes and 2,000 edges. The cost denoting total number of removing edges is set from 0 to 300 and the network without edges removed represents the initial network. Table 2 shows the configurations of the memetic algorithm parameters used on the optimization model.

Table 2 The parameters of the memetic algorithm.

| $X_n$ | $P_c$ | $P_l$ | μ | $P_m$ |
|---|---|---|---|---|
| 20 | 0.8 | 5/C | 0.7 | 500 |

Figure 4 shows the simulation results of the memetic algorithm (MA) and edge-betweenness adaptive strategy ($B_EAS$) of identifying the vital edges for the BA network and Chinese ARN. Looking at both networks via MA, we can see that the robustness $R$ decreases and the cost $C$ increases when removing edges. It can also be noticed that the MA is significantly better than the $B_EAS$. The difference between the two methods is especially high. It is obvious that the critical edges in the network are not extremely related with the edge-betweenness. Moreover, the memetic algorithm works better in the Chinese ARN decreasing its robustness $R$ faster when a few edges are removed. That is because of the Chinese ARN is not a small-world network. And the Chinese ARN is vulnerable because of its small clustering coefficient and large average shortest path length.

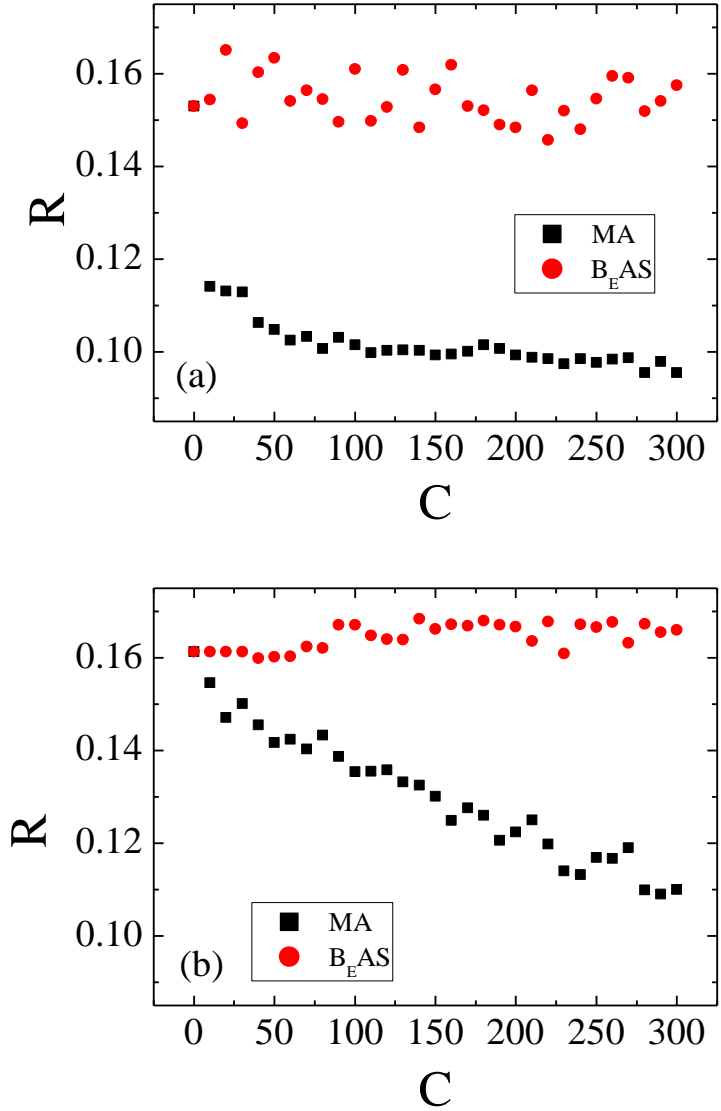

Fig. 4 The robustness $R$ as a function of the cost $C$ of removing edges for (a) the Chinese ARN and (b) the BA network.

We have seen that the memetic algorithm works on both Chinese ARN and BA network. In order to reveal the underlying mechanism clearly, we examine a toy model with a network containing 10 nodes and 17 edges (Figure 5). In figure 5, the blue lines are the existing edges and gray dotted lines are edges removed in that step. In the same way, the gray nodes mean that these nodes are removed and yellow nodes still exist. When no edge is removed from the network as shown in Fig. 5(a), the robustness of the initial network is 0.35. Here, three edges are removed to measure the criticality of these edges using the MA and $B_EAS$ methods. Since edges $e_{1,4}$, $e_{2,7}$ and $e_{4,9}$ have the highest edge-betweenness in the network, they are removed in the $B_EAS$ (Fig.5(b)) and now we have a new network named A. Similarly, edges $e_{1,3}$, $e_{4,9}$ and $e_{7,9}$ are identified as the most important in the MA, and we now have a new network B (Fig. 5(f)). For estimating which group of edges is critical, we compare the robustness of network A and B (see Fig. 5(c-e) and Fig. 5(g-i)). Figure 6 illustrates the corresponding solutions and the robustness of the solution for both methods.

In network A ($B_EAS$) we first remove node 2 with the highest-degree together with all edges connected with it: $e_{2,3}$, $e_{2,4}$, $e_{2,5}$, $e_{2,8}$ and $e_{2,10}$. $s(Q=1)$ of the new network is 0.9, which is the fraction of nodes in the giant component after removing 1 node (fig.5(c)). However in network B (MA), the value of *s(Q=1)* quickly decrease to 0.7 (fig.5(g)). Then, after the second nodes are removed, *s(Q=2)* in both networks A and B are 0.4 and 0.3 respectively (fig.5(d) and fig.(h)), which is reduced to 0.4 and 0.2 after the third node is removed (fig.5(e) and fig.5(i)). At the end, all nodes are removed from the network and the robustness of network A and B is 0.37 and 0.29 respectively. Thus, as previous results pointed, the critical edges in the network are not extremely related to the edge-betweenness, which apparently contradicts common intuitions.

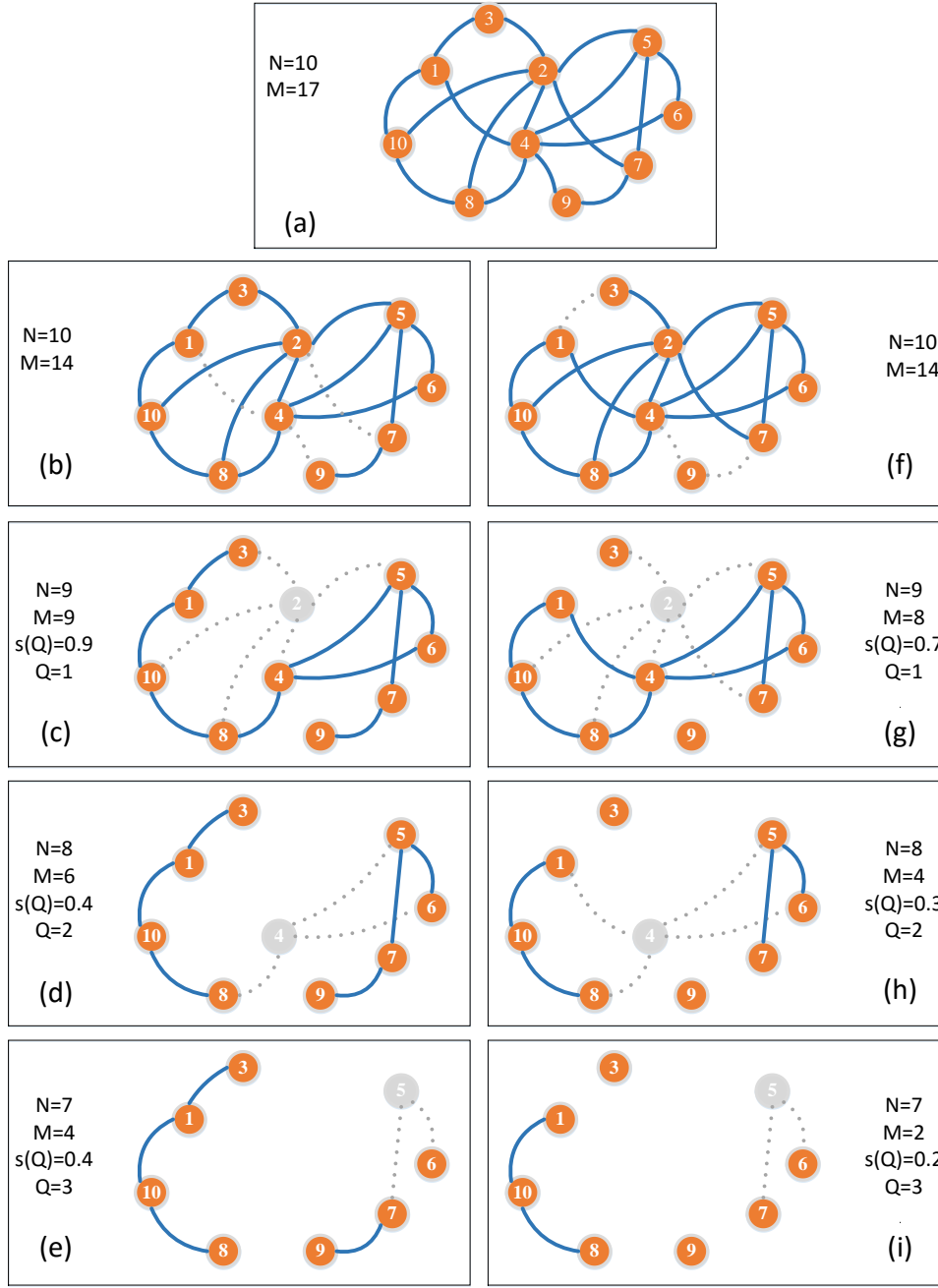

Fig.5 An illustration of malicious attack process of the toy model with 10 nodes via MA and strategy. Here in the network after removing *Q* nodes and the edges connected with them, *N* is the number of nodes, *M* is the number of edges, *s(Q)* is the fraction of nodes in largest component.

Table 3 illustrates the corresponding solutions and the robustness of the solution to both methods of the toy model in Fig.5. The results demonstrate that the most vital edges are not necessarily the edge with highest topological importance. Thus these edges identified by MA are important for the network robustness and should be protected to ensure the survivability of the network.

Table 3    An illustration of the vital edges identified by MA and $B_EAS$.

| $C$ | MA | $R$ | $B_E$AS | $R$ |
|---|---|---|---|---|
| 1 | $e_{1,10}$ | 0.33 | $e_{4,9}$ | 0.35 |
| 2 | $e_{4,6}, e_{5,6}$ | 0.32 | $e_{4,9}, e_{1,4}$ | 0.35 |
| 3 | $e_{1,3}, e_{4,9}, e_{7,9}$ | 0.29 | $e_{4,9}, e_{1,4}, e_{2,7}$ | 0.37 |
| 4 | $e_{2,5}, e_{2,7}, e_{7,9}, e_{8,10}$ | 0.27 | $e_{4,9}, e_{1,4}, e_{2,7}, e_{4,6}$ | 0.36 |
| 5 | $e_{1,10}, e_{4,9}, e_{5,7}, e_{7,9}, e_{8,10}$ | 0.24 | $e_{4,9}, e_{1,4}, e_{2,7}, e_{4,6}, e_{2,3}$ | 0.35 |
| 6 | $e_{1,3}, e_{1,10}, e_{2,3}, e_{2,5}, e_{2,7}, e_{7,9}$ | 0.24 | $e_{4,9}, e_{1,4}, e_{2,7}, e_{4,6}, e_{2,3}, e_{4,8}$ | 0.34 |

## 5. Conclusion

It is of great importance to improve the robustness of real networks. In this paper, we identified the vital edges in Chinese air route network, which lead to fast breakdown after targeted attacks. Our results revealed that the edge-betweenness, an index to measure the importance of edges in short paths, is of little relevance to this problem. Furthermore, we demonstrated that the memetic algorithm is able to pinpoint the edges that have been proven more important than edges of high edge-betweenness. We also confirmed these findings in scale-free model networks, hence offering novel insights of edge essentiality in various real networks. Thus, we think the vital edges identified by memetic algorithm should be especially protected to ensure a good performance of a network. In Chinese ARN this means that air traffic managers should foresight complex solutions when considering the closure of one vital air route segment.

## Acknowledgement

This paper is supported by the National Natural Science Foundation of China (Grant Nos. 91538204, 61425014, 61521091), National Key Research and Development Program of China (Grant No. 2016YFB1200100), and National Key Technology R&D Program of China (Grant No. 2015BAG15B01).